\documentclass[prl,showpacs,twocolumn,superscriptaddress]{revtex4}

\newcommand{\ket}[1]{|{#1}\rangle}
\newcommand{\bra}[1]{\langle{#1}|}

\newcommand{\be}{\begin{equation}}
\newcommand{\ee}{\end{equation}}

\usepackage{amsfonts}
\usepackage{amsmath}
\usepackage{rotating}
\usepackage{graphicx}
\usepackage{amssymb}
\usepackage{amsmath}
\usepackage{amssymb}
\usepackage{graphicx}
\usepackage{lscape}
\usepackage{color}
\usepackage{epstopdf}

\begin{document}
\title{Low-energy behaviour of strongly-interacting bosons on a flat-banded lattice above the critical filling factor}

\author{L.~G. Phillips}
\affiliation{SUPA, Institute of Photonics and Quantum Sciences, Heriot-Watt University, Edinburgh EH14 4AS, United Kingdom}
\author{G. De Chiara}
\affiliation{Centre for Theoretical Atomic, Molecular and Optical Physics,
School of Mathematics and Physics, Queen's University, Belfast BT7 1NN, United Kingdom}
\author{P. \"Ohberg}
\affiliation{SUPA, Institute of Photonics and Quantum Sciences, Heriot-Watt University, Edinburgh EH14 4AS, United Kingdom}
\author{M. Valiente}
\affiliation{SUPA, Institute of Photonics and Quantum Sciences, Heriot-Watt University, Edinburgh EH14 4AS, United Kingdom}
\begin{abstract}
Bosons interacting repulsively on a lattice with a flat lowest band energy dispersion may, at sufficiently small filling factors, enter into a Wigner-crystal-like phase. This phase is a consequence of the dispersionless nature of the system, which in turn implies the occurrence of single-particle localised eigenstates. We investigate one of these systems -- the sawtooth lattice -- filled with strongly repulsive bosons at filling factors infinitesimally above the critical point where the crystal phase is no longer the ground state. We find, in the hard-core limit, that the crystal retains its structure in all but one of its cells, where it is broken. The broken cell corresponds to an exotic kind of repulsively bound state, which becomes delocalised. We investigate the excitation spectrum of the system analytically and find that the bound state behaves as a single particle hopping on an effective lattice with reduced periodicity, and is therefore gapless. Thus, the addition of a single particle to a flat-banded system at critical filling is found to be enough to make kinetic behaviour manifest.

\end{abstract}
\maketitle
\paragraph{Introduction.}
Flat-banded lattices, that is, lattices with a large degenerate subspace of single particle solutions, have been the subject of interest for some time. For instance, they play a key role in the theory of ferromagnetism, where rigorous results by Lieb \cite{11}, Mielke \cite{12} and Tasaki \cite{10} guarantee the occurrence of ferromagnetism in flat-banded Hubbard models, without the need for unrealistic long range hopping terms. Also, the analogy between flat bands and the Landau levels enables the use of ultracold atomic systems \cite{9,13} as a means of experimenting with quantum Hall physics \cite{1,2,3}. The interface between flat-band ferromagnetism and topological band theory has also been studied \cite{16,17,18}. The above examples pertain to fermionic systems, which are the main target of study in condensed matter physics. On the other hand, it is possible to engineer flat-banded lattices for ultracold \emph{bosons} by loading bosonic atoms into optical lattices \cite{7,9}. Such systems are interesting in their own right, as they can be expected to support novel phases of matter not necessarily related to the quantum Hall effect or any other paradigmatic condensed matter phenomenon \cite{4,14}.

A flat band is simply an energy band in which the energy is constant, i.e. independent of the particle's momentum. In a flat band, kinetic energy is an irrelevancy and behaviour is governed entirely by interactions, so that even weakly interacting particles in the low-density limit enter a state that is strongly correlated and profoundly nonperturbative. Often, a consequence of such prepotency of interactions over kinetic terms is a Wigner-crystal-like ground state, in which the particles occupy non-overlapping localised eigenstates \cite{10,5,6,15}. In the repulsively interacting regime, and when the flat band is the band of lowest energy, this behaviour can be explained via a simple energetic argument. It is energetically unfavourable for particles to overlap, but occupying a superposition of orthogonal flat band modes which is zero over all but a few lattice sites incurs no energy penalty. The system can avoid the energy cost of double and higher occupancies by filling the lattice with non-overlapping localised eigenstates, and in this manner a crystal is formed. This picture, however, only holds true at low density. Above a critical filling factor $\nu_c$ ($\nu=N/L$ where $N$ is the particle number and $L$ is the number of lattice sites) there is insufficient space for every particle to occupy a localised state without any overlap, and the pure crystalline structure must be (at least partially) destroyed. The behaviour of such lattice models at slightly above $\nu_c$ has been studied recently by Huber and co-workers in \cite{4,8}, and by M\"oller and Cooper \cite{15}. In these works, the authors treat the weak-coupling limit, with the band gap much larger than the on-site interaction. They therefore assume that the ground state can be constructed entirely from (a projection onto) flat-band modes: an entirely justifiable approach, which provides excellent agreement with full-blown numerical calculations \cite{4}. However, if the interaction energy is much larger than the band gap, the particles cannot all be expected to stay in superpositions of flat band modes as in the weak-coupling regime, and it is unclear how states which have contributions from the upper bands enter the problem, and how kinetic behaviour, if at all, manifests. 

In this Letter, we investigate strongly-interacting bosons on a lattice supporting a flat lowest band. Specifically, we study the particularly simple sawtooth lattice (see Fig. 1), whose behaviour  in the weakly-interacting case has been the subject of previous works \cite{4}. By investigating the situation where the filling fraction is $\nu = \nu_c+\epsilon$, with $\epsilon = O(1/N)$, we find that kinetic behaviour does indeed occur at a filling slightly above the critical value, in the following, and rather unexpected, way. A two-body bound state is formed in the hard-core limit: a surprising result, given that the bosons comprising it do not overlap with each other. This bound state traverses the crystal as if it were a single particle acting under a pure hopping Hamiltonian, and moves with a quadratic dispersion relation at low energies, in stark contrast to the situation at and below critical filling where kinetic energy is completely quenched. Repulsively bound pairs in the Hubbard model have been studied \cite{22,23,24,25,26} and observed with ultracold atoms in optical lattices \cite{19} and nonlinear optical systems \cite{20}, but the physical situations treated in those studies and experiments are completely different from the scenario analysed here. For instance, in the above works the repulsively bound pairs exhibit large double occupancies for strong on-site interactions. By contrast, our result shows that it is possible to find repulsively bound pairs, \emph{in the medium}, even when the particles are completely forbidden from overlapping.
\begin{figure}[t]
\centering
\begin{turn}{270}
\includegraphics[width=0.1\textwidth]{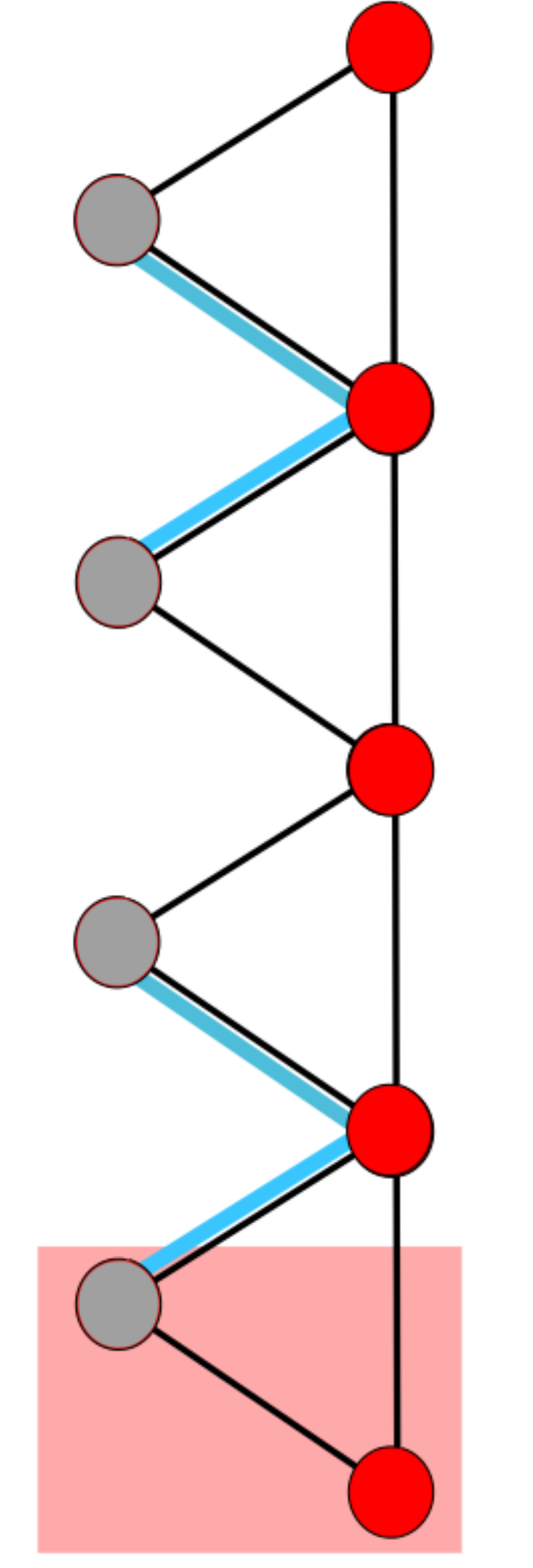}
\end{turn}
\caption{The sawtooth lattice, with a unit cell highlighted. A flat band occurs when the hopping amplitude between red sites is 1, and from red to grey is $\sqrt{2}$. The localised states associated with this band are shown in blue.}
\end{figure}
\paragraph{Sawtooth lattice below critical filling.}
The sawtooth lattice is effectively one dimensional (1D), being essentially a 1D chain with nearest and next-nearest-neighbour hopping. For simplicity, and because we have in mind bosonic atoms in an optical lattice for the experimental realisation, we model the system's dynamics by the Bose-Hubbard Hamiltonian with on-site interaction $U$,
\begin{equation}
H=\sum_{i,j}t_{i,j}b^{\dagger}_ib_j+\frac{U}{2}\sum_in_i(n_i-1),
\label{2qh}
\end{equation}
where $b_i$ ($b_i^{\dagger}$) is the bosonic annihilation (creation) operator at site $i$, $n_i=b_i^{\dagger}b_i$ is the number operator at site $i$, and where $t_{ij}$ are hopping constants, given by  
\begin{align}
t_{2m,j}&=t\left(\sqrt{2}\delta_{|2m-j|,2}+\delta_{|2m-j|,1}\right),\label{kin1}\\
t_{2m+1,j}&=t\delta_{|2m+1-j|,1},\label{kin2}
\end{align}
where $m$ is an integer, and $t>0$ is the nearest-neighbour tunneling rate. Note that we assume periodic boundary conditions, an even number of lattice sites, and have set the lattice constant to unity. To confirm that our choices for the $t_{ij}$ do indeed give rise to a flat band, the single particle problem must be solved. One may pass to first quantisation and write the stationary Schr\"odinger equation $H\psi=E\psi$ as:
\begin{equation} 
\sum_{\mu=\pm1}\left[\sqrt{2}\psi(j+\mu)+\frac{\left(1+(-1)^j\right)}{2}\psi(j+2\mu)\right]=E\psi(j). \label{1qh}
\end{equation}
Using Bloch's theorem to write the wavefunction as $\psi_k(j)=\phi_k(j)e^{ikj}$, where the $\phi_k(x)$ are functions of periodicity 2, gives two coupled equations,
\begin{align} 
E\phi_k(1)&=2t\sqrt{2}\cos{2k}\,\phi_k(0) \label{1psy}\\
(E-2t\cos{2k})\phi_k(0)&=2t\sqrt{2}\cos{k}\,\phi_k(1)\label{1psy2} .
\end{align}
This system is easily solved for the energy $E$, revealing the lowest flat and the excited dispersive bands,
\begin{align}
E_0(k)&=-2t,\\
E_1(k)&= 2t\left(1+\cos{2k}\right).\label{E1k}
\end{align} 
The (unnormalized) localised eigenstates associated with the flat band (see \cite{10} for mathematical details on the relationship between flat bands and localised states) are given by 
\begin{equation} 
V^{\dagger}_i|0\rangle=(\sqrt{2}b^{\dagger}_{2i}-b^{\dagger}_{2i+1}-b^{\dagger}_{2i-1})|0\rangle.
\end{equation}
It is easy to check that $HV_i^{\dagger}|0\rangle=E_0V_i^{\dagger}|0\rangle$, from which one immediately concludes that the $V_i^{\dagger}|0\rangle$ are indeed superpositions of orthogonal flat band modes. Clearly, at most $L/4$ of these states can fit on the lattice without overlapping. Thus, up to $\nu=\nu_c=1/4$, the (not necessarily orthogonal) degenerate many body ground states take the form
\begin{equation}
|\psi_0\rangle=\prod V^{\dagger}_i|0\rangle
\end{equation}
where the product is over a set of $N$ integers $\{i_1,i_2,...,i_N: |i_i-i_j|>1 \,\forall i,j \}$. In what follows we take the hard-core limit $U\to\infty$, and thus allow at most one particle per site.    
\paragraph{A trial wavefunction.}
We now attempt to treat the sawtooth lattice at a single particle above critical filling. There are two ways of doing so without changing the periodic properties of the system: one may either add one particle on top of the preexisting $N=L/4$ (equivalent to reducing the size of the lattice by four sites), or remove one unit cell (two sites) from the lattice. We choose the latter option which is the simpler since, in the language of \cite{4}, it creates a single domain wall, as opposed to the former, which creates two. We have already pointed out the inadequacy of perturbative methods, so we adopt a variational approach. Thus our primary task is to decide upon a sensible ansatz. To this end, note that there are two ways in which the crystal phase might be destroyed: $i)$ the extra particle may become delocalised and upset the structure of the entire crystal, or $ii)$ it may remain localised and break one or several cells of the crystal, leaving the rest intact. We were able to decide between these two scenarios with the help of numerical evidence from exact diagonalization (ED) with up to 5 particles and periodic boundary conditions and from density-matrix-renormalization group (DMRG) \cite{White,Chiara} with up to 25 particles, with open boundary conditions. Both calculations show that the ground state energy scales with particle number as $E(N)=(N-2)E_0+C$, with $2E_0<C<E_0$ . This strongly suggests that the extra particle breaks the crystal in a single cell, leaving $N-2$ of the localised states intact (in scenario $i$ the energy would scale as something like $E(N)=NC'$).
\begin{figure}
\centering
\includegraphics[width=0.5\textwidth]{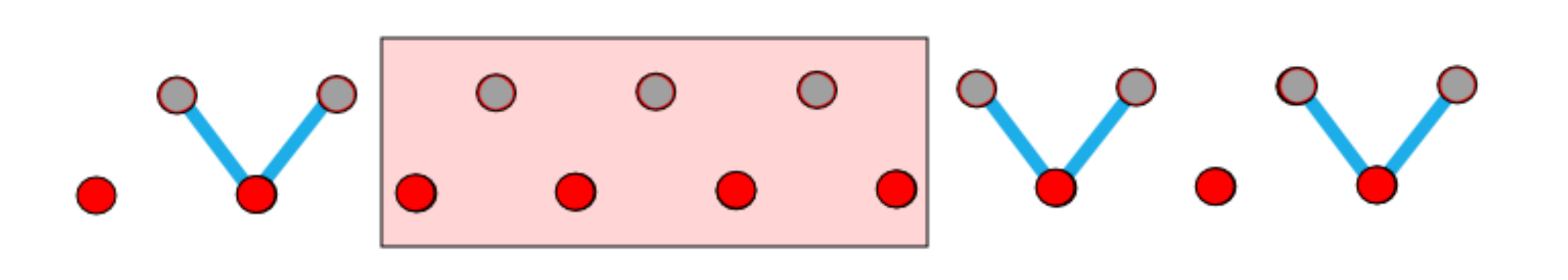}
\caption{A pictorical representation of a component of the ansatz. The localised states are shown in blue. The highlighted block contains two particles and is diagonalised numerically. The full ansatz is a superposition of states like this, with the highlighted block starting on each red site.} 
\end{figure}

As we noted above, at one particle above critical filling the lattice may be either one or two unit cells too small to accomodate the crystal. If it is one unit cell too small, the structural disruption will be confined to a seven-site block, which will contain two particles. Denote a state in which the disrupted block begins on the $2i^{th}$ site as $|\psi_i\rangle$ (see Fig. 2), so
\begin{equation}
|\psi_i\rangle=B^{\dagger}_i\prod_{l=1}^{N-2}V^{\dagger}_{i+2l+2}|0\rangle,
\end{equation}
where $B^{\dagger}_i =  \sum_{j=0}^{5}\sum_{k=j+1}^{6}\alpha_{jk}b^{\dagger}_{2i+j}b^{\dagger}_{2i+k}$. The $\alpha_{jk}$ are chosen so that $B^{\dagger}_i|0\rangle$ is the ground state of a system of two particles in seven sites with open boundary conditions. Hence, we can write
\begin{equation}
HB^{\dagger}_i|0\rangle=E_B B^{\dagger}_i|0\rangle+X_i^{\dagger}|0\rangle,
\end{equation}
where $E_B$ is the seven-site ground state energy and $X^{\dagger}_i$ creates the terms that ``leak" out from the disrupted block when the Hamiltonian is applied. 
Because of translational invariance, no particular block can be expected to contain the two-body state. Accordingly, eigenstates should be superpositions of the $|\psi_i\rangle$: $|\Psi\rangle=\sum_i\beta_i|\psi_i\rangle$. This last is our ansatz, with which we seek to minimize the energy expectation value, using the (complex-valued) $\beta_i$ as variational parameters. We must solve
\begin{equation}
\frac{\delta}{\delta \beta^*_i}\left(\langle\Psi|H|\Psi\rangle-E\langle\Psi|\Psi\rangle\right)=0,\label{variationalthingy}
\end{equation}
where $E$ is a Lagrange multiplier to be identified with the variational energies. After simple manipulation, Eq. (\ref{variationalthingy}) becomes
\begin{equation}\label{eq:gev}
(C-E_B)\sum_j\langle\psi_i|\psi_j\rangle\beta_j=\sum_j\langle\psi_i|X^{\dagger}_j|0\rangle\beta_j
\end{equation}
\begin{figure}[t]
\centering
\includegraphics[width=0.5\textwidth]{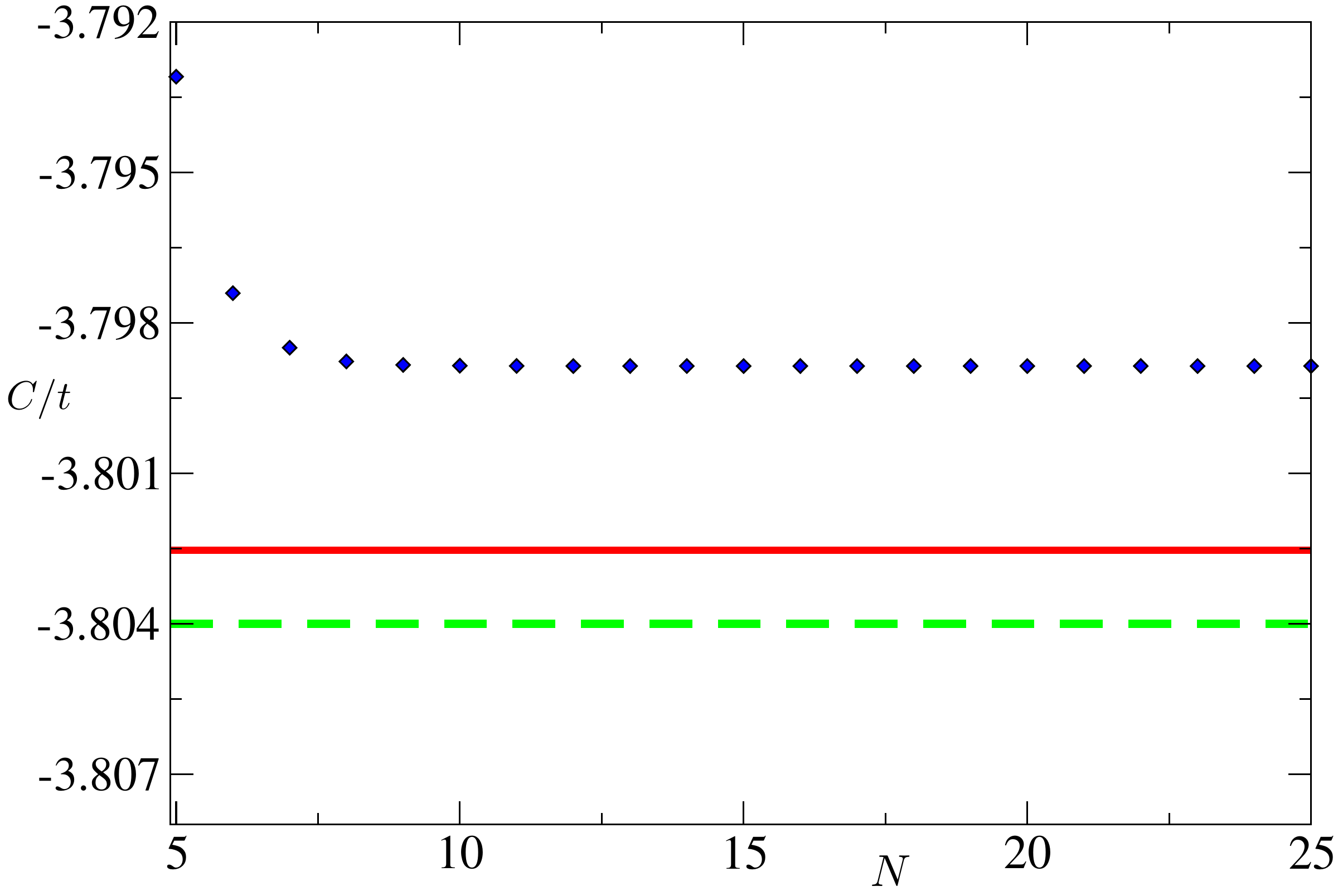}
\caption{Comparison of energy obtained via minimisation for various particle numbers (blue diamonds) with DMRG energy for 25 particles (solid red line) and energy from exact diagonalisation for 5 particles (dashed green line). As particle number is increased, the energy obtained via minimisation quickly tends to within $0.005E_0$ of the DMRG value.}
\end{figure}
with $C=E-(N-2)E_0$. Notice that, since the states $\ket{\psi_i}$ are not orthogonal to each other, the above equation represents a generalized eigenvalue problem (GEP). The lowest value of $C$ obtained by solving this GEP numerically for 25 particles agrees with the ground state value obtained via DMRG to within $0.5\%$ of the characteristic energy $E_0$, which confirms that our ansatz is indeed a sensible one, and that the solutions of Eq. (\ref{eq:gev}) furnish a good approximation to the set of exact eigenstates. See Fig. 3 for a comparison of the $C$ obtained from functional minimization with that from DMRG \footnote{$C$, rather than $E$, is the pertinent quantity when it comes to assessing the accuracy of results here, since for any large value of $N$, the breakage energy, being of $\mathcal{O}(1)$, will be washed out by the trivial contribution of $\mathcal{O}(N)$ from the localised states.}.
\paragraph{Results and discussion.}
Of course, Eq. (\ref{eq:gev}) has $L/2$ solutions. The lowest energy solution is unique, and each subsequent solution is twofold degenerate, suggesting the existence of a quasimomentum-like quantum number.
Each solution yields a set of $\beta_j$. Acting on our intuition about the quasimomentum, we label each set by an integer $n$, and have the energy increase monotonically with $|n|$. We let $n$ run from $-L/4$ to $L/4-1$. The degenerate states are labelled $n=\pm|n|$, and the unique ground state has $n=0$. With this labelling scheme, if $k$ is defined as $k=2\pi n/L$, we have verified numerically that $\beta_j^{(n)}=(-1)^{jn}e^{ijk}$ to machine accuracy, so
\begin{equation}
|\Psi_n\rangle=\sum_j(-1)^{jn}e^{ijk}|\psi_j\rangle.
\end{equation}
\begin{figure}[t]\label{disp}
\centering
\includegraphics[width=0.5\textwidth]{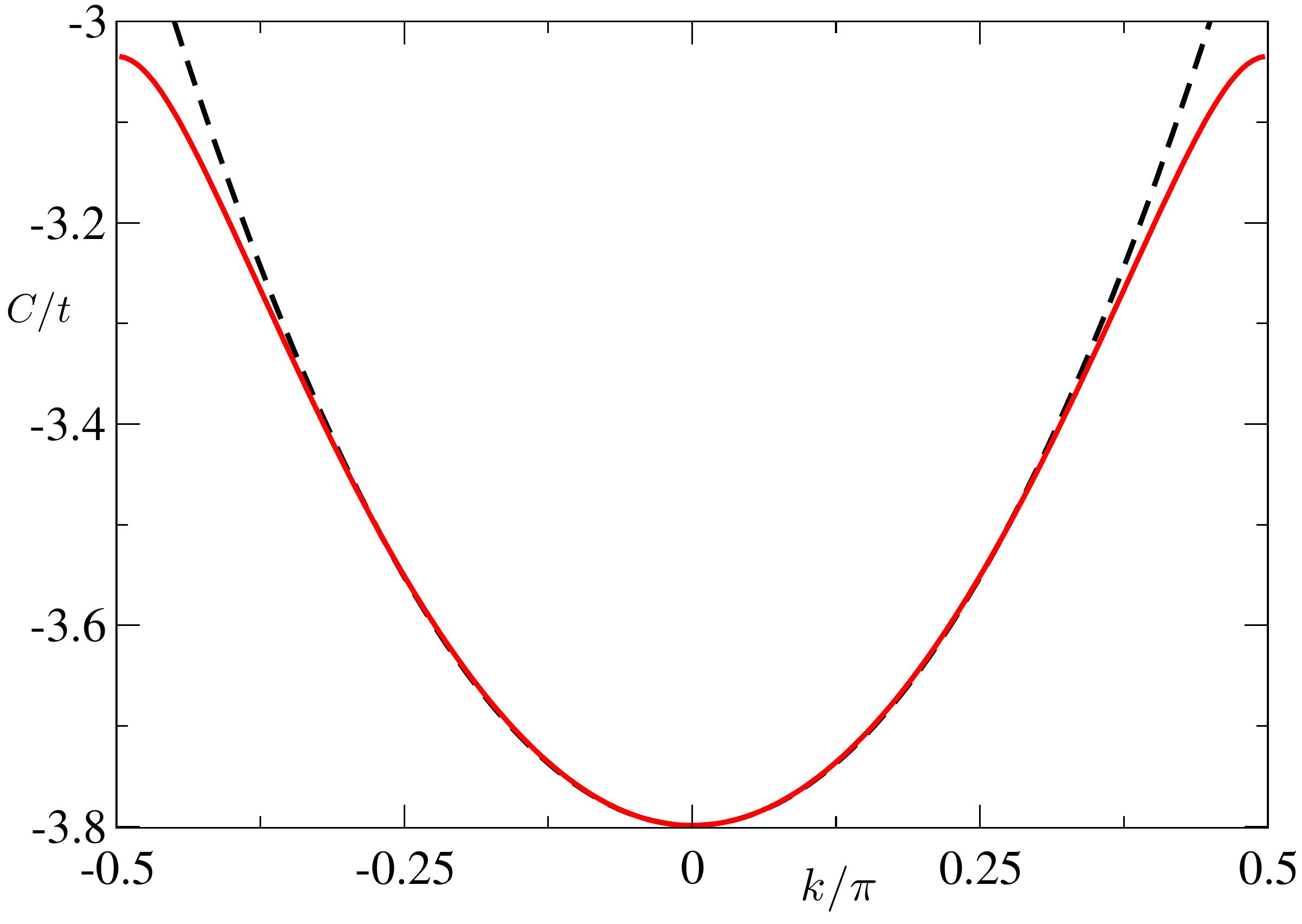}
\caption{Dispersion relation for the moving bound state (solid red line), together with a quadratic function representing the dispersion of a free particle of mass $m^*/t=1.25$ (dashed black line)}
\end{figure}
The solution is equivalent to that of a single particle hopping on a lattice of periodicity 2, with Bloch functions $\phi_n(j)=(-1)^{jn}$, and the repulsively bound state playing the role of the particle. The dispersion relation is plotted in Fig. 4. It is exactly quadratic in the low energy sector, with an effective mass of approximately $m^*/t= 1.25$. The effective mass is apparently very large, as it is $\sim 10$ times higher than the single-particle effective mass in the dispersive band $E_1(k)$, Eq. (\ref{E1k}). However, if we compare this to the effective mass of the excitations in the weak-coupling limit, which is of $\mathcal{O}(t^2/U)\to \infty$ \cite{4}, we find that the effective mass is exceptionally low and therefore the contribution from the excited band is highly relevant. At low energies, then, there is a close analogy between our system at $N=N_c+1$ and a (heavy) single free particle in the continuum.

Our predictions can be verified experimentally by measureming the ground state momentum distribution, an experiment that is routinely performed with ultracold atoms in optical lattices \cite{21,Bloch,27}. We now calculate the expected results, and while doing so demonstrate non-analytic behaviour -- an instability -- around critical filling. At $\nu\le \nu_c$, it is a simple matter to show that
\be
\langle n_k \rangle_{\nu\le \nu_c}\equiv \bra{\psi_0} n_k\ket{\psi_0} = \frac{\nu}{4}\left(\sqrt{2}-\cos{k}\right)^2.\label{momdist1}
\ee
At $\nu=\nu_c+\epsilon$ the momentum density deviates from eqn. (\ref{momdist1}) slightly. This deviation is due to the addition of a single particle and hence rather small, so a direct measurement of $\langle n_k\rangle_{\nu_c}$ is unlikely to give useable data. Rather, measuring $\langle n_k\rangle_{\nu_c\pm\epsilon}$ and $\langle n_k\rangle_{\nu_c}$, and thence calculating the right derivative,
\be
\frac{\langle n_k\rangle_{\nu_c+\epsilon}-\langle n_k\rangle_{\nu_c}}{\epsilon}=\left.\frac{\partial\langle n_k\rangle}{\partial \nu}\right|_{\nu_c^+}+\mathcal{O}(\epsilon),
\ee
would yield data that can be meaningfully compared with the derivative obtained from our model, shown in Fig. 5. It is clear from eqn. (\ref{momdist1}) and Fig. 5 that the right and left derivatives do not agree at $\nu_c$; this singularity is a signature of the destruction of the crystalline structure. 
To further support our conclusions we compare the derivative obtained from our model with results from exact diagonalisation, and find good agreement. 

In summary, although kinetic energy is quenched at $\nu_c$ or below, we find that an extra particle above $\nu_c$ does away with this quenching: the interaction is no longer the only relevent parameter. Kinetic behavour manifests in the form of a novel repulsively bound pair travelling through the lattice. The emergence of kinetic behaviour and the existance of this non-overlapping repulsively bound state are our main findings.
The fact that the excitation spectrum is gapless leads us to believe that we have found the lowest lying states, and this, together with the closeness between our groundstate energy and the DMRG result and the agreement on the momentum distribution between exact diagonalisation and our model, suggests that we have captured all the essential low-energy physics with our picture.
\paragraph{Future work and outlook.}
The single-particle-like nature of solutions at $N=N_c+1$ indicates to us the possibilty of modelling behaviour at a few particles above $N_c$ via a theory (perhaps exactly solvable) of interacting bound states. Further, we suspect that our findings are not limited in their applicability to the sawtooth lattice, and give insight into the general nature of the destruction of lattice Wigner-like crystals by overfilling. A confirmation or refutation of this suspicion would be interesting; were it to be confirmed, we would have a general prescription for treating flat-banded lattice models above $\nu_c$.
\begin{figure}
\centering
\includegraphics[width=0.5\textwidth]{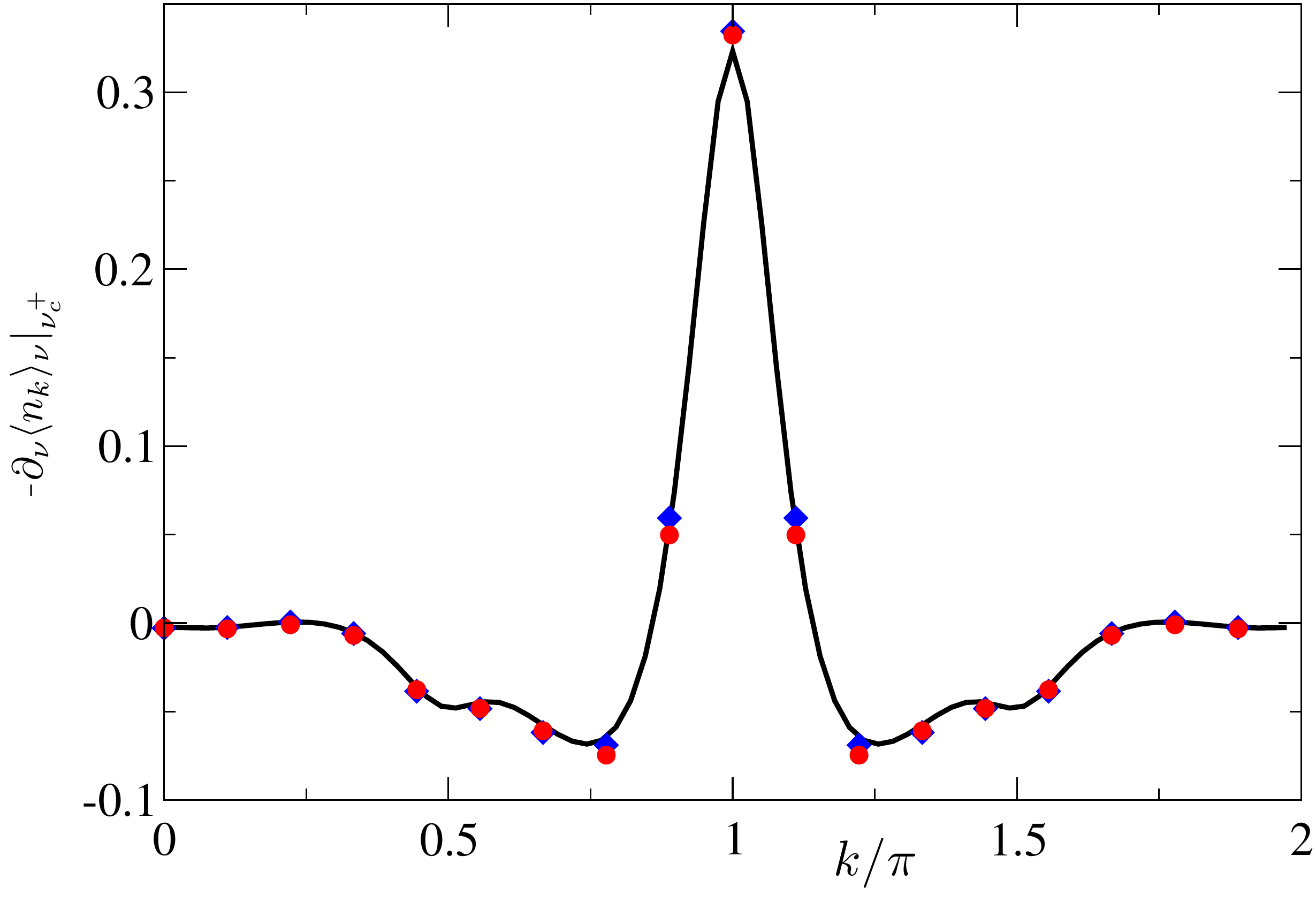}
\caption{\emph{Minus} the right derivative of momentum density as a fucntion of filling fraction at critical filling, as per eqn. (17), as obtianed from our ansatz with 20 particles (black line), 5 particles (blue diamonds) and from exact diagonalisation with 5 particles (red circles).} 
\label{fig5}
\end{figure}
\acknowledgements{We would like to thank S.~D. Huber for useful discussions. L.G.P. acknowledges support from the EPSRC CM-DTC, P.\"O. and M.V.  acknowledge support from EPSRC grant No. EP/J001392/1, G.D.C.acknowledges support from the UK EPSRC (EP/L005026/1 and EP/K029371/1), the John Templeton Foundation (grant ID 43467), and the EU Collaborative Project TherMiQ (Grant Agreement 618074).}

\bibliographystyle{unsrt}

\begin{thebibliography}{99}
\bibitem{11} E. H. Lieb,
Phys. Rev. Lett. \textbf{62}, 1201 (1989).
\bibitem{12} A. Mielke,
Phys. Lett. \textbf{A174}, 443 (1993).
\bibitem{10} H. Tasaki,
Prog. Theor. Phys. \textbf{99}, 489 (1998).
\bibitem{9} G-B. Jo, J. Guzman, C.~K. Thomas, P. Hosur, A. Vishwanath, D.~M. Stamper-Kurn,
Phy. Rev. Lett. \textbf{108}, 045305 (2012).
\bibitem{13} L. Mazza, A. Bermudez, N. Goldman, M. Rizzi, M.~A. Martin-Delgado, M. Lewenstein,
New J. Phys. \textbf{14} 015007 (2012).
\bibitem{1} K. Sun, Z.~G. Gu, H. Katsura and S. Das Sarma,
Phys. Rev. Lett. \textbf{106}, 236803 (2011).
\bibitem{2} Y.~F. Wang , Z.~G. Gu, C.~D. Gong and D.~N. Sheng,
Phys. Rev. Lett. \textbf{107}, 146803 (2011).
\bibitem{3} D.~N. Sheng, Z.~G. Gu, K. Sun and L. Sheng,
Nature Comm. \textbf{2}, 389 (2011).
\bibitem{16} H. Katsura, I. Maruyama, A. Tanaka, H. Tasaki,
E.P.L. \textbf{91}, 54007 (2010).
\bibitem{17} J. He, B. Wang, S-P. Kou,
Phys. Rev. B \textbf{86}, 235146 (2012).
\bibitem{18} T. Paananen, H. Gerber, M. G\"otte, T. Dahm,
New J. Phys. \textbf{16} 033019;
\bibitem{7} G. Ritt, C. Geckeler, T. Salger, G. Cennini, M. Weitz,
Phys. Rev. A \textbf{74}, 063622 (2006).
\bibitem{4} S.~D. Huber and E. Altman,
Phys. Rev. B \textbf{82}, 184502 (2010).
\bibitem{14} M. Hyrk\"as, V. Apaja, M. Manninen,
Phys. Rev. A \textbf{87} 023614 (2013).
\bibitem{Wigner} E. Wigner,
Phys. Rev. \textbf{46} (11) 1102 (1934).
\bibitem{5} J.~T. Chalker, T.~S. Pickles and P. Shukla,
Phys. Rev. B \textbf{82}, 104209 (2010).
\bibitem{6} R. Takahashi and S. Murakami,
Phys. Rev. B \textbf{88}, 235303 (2013).
\bibitem{15} G. M\"oller, N.~R. Cooper,
Phys. Rev. Lett. \textbf{108} 045306 (2012).
\bibitem{8} M. Tovmasyan, E.~P.~L. van Nieywenburg, S.~D. Huber,
Phys. Rev. B \textbf{88}, 220510 (2013).
Nature \textbf{429} 277-81 (2004).
\bibitem{22} M. Valiente, D. Petrosyan,
J. Phys. B: At. Mol. Opt. Phys. \textbf{41} 161002 (2008).
\bibitem{23} M. Valiente, D. Petrosyan,
J. Phys. B: At. Mol. Opt. Phys. \textbf{42} 121001 (2009).
\bibitem{24} M. Valiente,
Phys. Rev. A \textbf{81}, 042102 (2010).
\bibitem{25} P. Pill, K. M{\o}lmer,
Phys. Rev. A \textbf{76}, 023607 (2007).
\bibitem{26} J.~C. Sanders, O. Odong, J. Javanainen, M. Mackie,
Phys. Rev. A \textbf{83}, 031607 (2011).
\bibitem{19} K. Winkler, G. Thalhammer, F. Lang, R. Grimm, J. Hecker Denschlag, A.~J. Daley, A. Kantian, H.~P. B\"uchler, P. Zoller,
Nature \textbf{441}, 853-856 (2006).
\bibitem{20} Y. Lahini, M. Verbin, S. D. Huber, Y. Bromberg, R. Pugatch, Y. Silberg,
Phys. Rev. A \textbf{86} 011603 (2012).
\bibitem{White} S.R. White,
Phys. Rev. Lett. \textbf{69} 2863 (1992).
\bibitem{Chiara} G. De Chiara, L. Lepori, M. Lewenstein, A. Sanpera,
Comp. Theor. Nanos. \textbf{5} 1277 (2008).
\bibitem{21} B. Paredes, A. Widera, V. Murg, O. Mandel, S. F\"olling, I. Cirac, G.~V. Shlyapnikov, T.~W. H\"ansch, I. Bloch,
\bibitem{Bloch} Immanuel Bloch, Jean Dalibard, Wilhelm Zwerger,
Rev. Mod. Phys. \textbf{80}, 885 (2008).
\bibitem{27} P. Longo, J. Evers,
Phys. Rev. Lett. \textbf{112}, 193601 (2014).


\end{thebibliography}

\end{document}